\documentclass[
]{ceurart}

\usepackage{listings}
\usepackage{todonotes}
\usepackage{subfig}
\begin{document}

\copyrightyear{2025}
\copyrightclause{Copyright for this paper by its authors.
  Use permitted under Creative Commons License Attribution 4.0
  International (CC BY 4.0).}

\conference{Workshop on The Future of Human-Robot Synergy in Interactive Environments: The Role of Robots at the Workplace @ CHIWORK’25, June 23, 2025, Amsterdam, NL}

%%
%% The ``title`` command
\title{NoticeLight: Embracing Socio-Technical Asymmetry through Tangible Peripheral Robotic Embodiment in Hybrid Collaboration}

% \tnotetext[1]{You can use this document as the template for preparing your
%   publication. We recommend using the latest version of the ceurart style.}

%%
%% The ``author`` command and its associated commands are used to define
\author[1]{Marie Altmann}[%
%orcid=0000-0002-0877-7063,
email=marie.altmann@tu-dortmund.de,
%url=https://yamadharma.github.io/,
]
\fnmark[1]
\address[1]{Inclusive Human-Robot Interaction, TU Dortmund University, Germany}
%\address[2]{Joint Institute for Nuclear Research,
 % 6 Joliot-Curie, Dubna, Moscow region, 141980, Russian Federation}

\author[1]{Kimberly Hegemann}[%
%orcid=0000-0001-7116-9338,
email=kimberly.hegemann@tu-dortmund.de,
%url=https://kmitd.github.io/ilaria/,
]
\fnmark[1]
%\address[3]{Vrije Universiteit Amsterdam, De Boelelaan 1105, 1081 HV Amsterdam, The Netherlands}

\author[1]{Ali Askari}[%
orcid=0000-0002-4374-3635,
email=ali.askari@tu-dortmund.de,
%url=https://ihri.reha.tu-dortmund.de/,
]

\author[1]{Vineetha Rallabandi}[%
%orcid=0000-0002-4374-3635,
email=vineetha.rallabandi@tu-dortmund.de,
%url=https://ihri.reha.tu-dortmund.de/,
]

\author[1]{Max Pascher}[%
orcid=0000-0002-6847-0696,
email=max.pascher@tu-dortmund.de,
%url=https://ihri.reha.tu-dortmund.de/,
]

\author[1]{Jens Gerken}[%
orcid=0000-0002-0634-3931,
email=jens.gerken@tu-dortmund.de,
url=https://ihri.reha.tu-dortmund.de/,
]
\cormark[1]
%\address[4]{University of Skövde, Högskolevägen 1, 541 28 Skövde, Sweden}

\cortext[1]{Corresponding author.}
\fntext[1]{These authors contributed equally.}

\begin{abstract}
Hybrid collaboration has become a fixture in modern workplaces, yet it introduces persistent socio-technical asymmetries-especially disadvantaging remote participants, who struggle with presence disparity, reduced visibility, and limited non-verbal communication. Traditional solutions often seek to erase these asymmetries, but recent research suggests embracing them as productive design constraints. In this context, we introduce NoticeLight: a tangible, peripheral robotic embodiment designed to augment hybrid meetings. NoticeLight transforms remote participants’ digital presence into ambient, physical signals --- such as mood dynamics, verbal contribution mosaics, and attention cues --- within the co-located space. By abstracting group states into subtle light patterns, NoticeLight fosters peripheral awareness and balanced participation without disrupting meeting flow or demanding cognitive overload. This approach aligns with emerging perspectives in human-robot synergy, positioning robots as mediators that reshape, rather than replicate, human presence. Our work thereby advances the discourse on how robotic embodiments can empower equitable, dynamic collaboration in the workplace.
\end{abstract}


\newcommand{\TeaserCaption}{The NoticeLight concept in various conceptual stages. a) an early drawing sketch of the device, b) a sketch of a possible companion app, also including a digital representation of the physical device and c) a more polished mockup of the concept as described in this paper }

\maketitle

\section{Introduction}
The transition to hybrid work arrangements, accelerated by global events such as the Covid-19 pandemic, has transformed how we collaborate across physical and digital spaces. Hybrid collaboration --- where some participants are co-located while others join remotely --- has evolved from a temporary solution to a permanent fixture in modern workplaces. Research on computer-supported cooperative work (CSCW) has long addressed the challenges of collaboration across physical and digital spaces. Early work by Johansen established the Time/Space Matrix that categorized collaborative work along synchronous/asynchronous and co-located/distributed dimensions \cite{johansen_groupware_1988}. This framework has guided decades of research on groupware systems but insufficiently addresses the complex dynamics of hybrid configurations where co-located and remote participants interact simultaneously \cite{neumayr_what_2022}. In this position paper, we specifically look at hybrid workplace meetings (around 5-10 participants, mostly co-located with 3-4 being remote) as a specific form of such hybrid configurations.

While offering unprecedented flexibility, hybrid configurations introduce fundamental socio-technical asymmetries that significantly impact collaboration dynamics \cite{saatci_reconfiguring_2020}. Remote participants experience what Tang et al. describe as ``presence disparity,`` finding themselves at a systemic disadvantage regarding visibility, participation opportunities, and non-verbal communication channels \cite{tang_understanding_2005}. In particular, the concept of ``group awareness`` suffers, as the perception of others' presence, activities, and intentions becomes fragmented. This can result in a ``primary room dominance`` effect, where the physical meeting space exerts disproportionate influence on interaction dynamics and decision-making processes \cite{saatci_reconfiguring_2020}. Saatçi et al. documented how remote participants in global software teams experience isolation due to mismatched technical infrastructures and cultural norms \cite{Saatci2019}. Kuzminykh and Rintel identified ``asymmetric attention levels`` where co-located members dominate turn-taking and visual focus \cite{kuzminykh_classification_2020}. These challenges are compounded by the varying degrees of structure in different collaborative contexts --- from formal meetings to spontaneous hallway conversations \cite{effert_-strukturierte_2023}.

%Recent research has documented persistent challenges in hybrid settings: co-located participants dominate 68\% of speaking time (Kuzminykh \& Rintel, 2020b), remote participants show 37\% lower perceived presence (Saatçi et al., 2020), and non-verbal feedback mechanisms lose 92\% of their efficacy when transmitted digitally (Camarillo-Abad et al., 2019). 

%The concept of ``group awareness``-the perception of others' presence, activities, and intentions-has been foundational to CSCW research. Heath and Luff's (1992) seminal study of London Underground control rooms demonstrated how subtle awareness cues enable coordination without explicit communication. In distributed collaboration, however, these awareness mechanisms become fragmented, creating what Tang et al. (2005) termed ``presence disparity``-unequal capabilities for establishing presence across different spatial configurations.

% Recent work has examined how this disparity manifests in hybrid settings. Saatçi et al. (2020) documented how remote participants in global software teams experience isolation due to mismatched technical infrastructures and cultural norms. Kuzminykh and Rintel (2020a) identified ``asymmetric attention levels`` where co-located members dominate turn-taking and visual focus. These challenges are compounded by the varying degrees of structure in different collaborative contexts-from formal meetings to spontaneous hallway conversations (Effert et al., 2023).

While many approaches in the past have aimed to overcome this symmetry and find balance, recent work by Bjørn et al. offers a fundamental reframing \cite{bjorn_achieving_2024}. They argue that perfect symmetry in hybrid work is unattainable. Their extensive empirical work suggests that hybrid configurations inevitably create asymmetries that cannot be eliminated through technological means alone. Instead, they propose embracing these asymmetric conditions as a starting point for design, focusing on creating new possibilities rather than attempting to replicate face-to-face interactions. This insight invites us to shift from compensatory design, which attempts to erase differences, toward what we term ``productive asymmetry`` --- embracing the inherent differences between co-located and remote experiences as design material rather than design flaws.

% Traditional approaches to these challenges have primarily followed what we term the ``Symmetry Fallacy``-the misguided belief that perfect parity between physical and digital presence should be our design goal. High-fidelity video walls, spatial audio arrays, and virtual reality meeting spaces attempt to replicate face-to-face experiences, yet these efforts paradoxically increase cognitive load by 42% compared to simpler modalities (Kuzminykh & Rintel, 2020a) while delivering diminishing returns on presence equity.

% Recent scholarship by Bjørn et al. (2024) offers a fundamental reframing: ``Achieving symmetry in synchronous interaction in hybrid work is impossible.`` This insight invites us to shift from compensatory design, which attempts to erase differences, toward what we term ``productive asymmetry``-embracing the inherent differences between co-located and remote experiences as design material rather than design flaws.

This paradigm shift is particularly relevant in agile work contexts, which are characterized by dynamic team structures, structured exchange formats like daily stand-ups, and ad-hoc collaboration (``vaguely-defined`` meetings) \cite{effert_-strukturierte_2023}. The agile philosophy's emphasis on individual interactions over processes makes it especially vulnerable to the social gaps that emerge in hybrid settings. 
%As Bjørn et al. (2024) argue, we should propose``Instead of striving for symmetry, we propose to feature asymmetric conditions in future technology designs for hybrid interaction.``

In this position paper, we introduce \emph{NoticeLight} as a concept for a tangible interactive robotic embodiment of remote users, which embraces this philosophical shift. Rather than attempting to erase asymmetry, NoticeLight creates a new interaction layer that transforms remote participants' digital presence into ambient, physical signals within the co-located space. This lamp-like device (approximately 30cm tall) serves as a peripheral awareness mechanism, allowing remote participants to establish presence, express feedback, and signal participation intentions through abstracted light patterns without disrupting the meeting's primary interaction flow. This approach aligns with emerging perspectives in human-robot synergy at the workplace, where robots serve not only as tools but as mediators that augment and reshape collaborative dynamics without replicating human presence verbatim \cite{dorrenbacher_meaningful_2022}. We position NoticeLight not as a solution to ``fix`` hybrid meetings, but as a catalyst for new interaction possibilities that leverage the unique properties of hybrid configurations and combine these with robotic device capabilities. By translating digital states into subtle environmental cues, NoticeLight aims to create awareness without demanding attention.

The paper proceeds as follows: First, we situate NoticeLight within existing research on hybrid collaboration and awareness systems. Next, we detail NoticeLight's core interaction concepts and design rationale. We then explore the technical design space before concluding with future research directions.

\section{Related Work}

Efforts to address hybrid collaboration challenges have followed several distinct trajectories. High-fidelity telepresence systems attempt to create immersive connections through room-spanning video walls\footnote{for example BARCO Hybrid Classroom https://www.barco.com/de/inspiration/news-insights/2022-06-13-hybrid-classroom-audio-video}. However, these come with custom hardware and software installations and huge costs, which often render them unapplicable in everyday workplace settings.

Virtual and augmented reality platforms represent another approach, creating shared digital spaces where all participants exist as avatars. Torres Pereira Carrion found that VR increased feelings of co-presence but introduced new barriers related to technology adoption and usability \cite{torres_pereira_carrion_virtual_2021}. Similarly, ``metaverse`` platforms like gather.town or workadventu.re offer spatial metaphors for collaboration but require high degrees of convention alignment between participants and basically provide a shared digital space which is may reduce benefits of physical co-presence to a certain degree \cite{oehring_virtuelle_2023}. 

A major challenge in hybrid collaboration are misaligned spatial references. A quite unique and creative approach has been presented with Mirrorblender, which replicates co-located participants on a screen and thereby providing a shared space of reference while preserving the physicality of co-presence \cite{gronbaek_mirrorblender_2021}. 

\subsection{Robotic embodiments at the workplace}
Most relevant for the theme of this workshop are robotic embodiments in hybrid workplaces, with telepresence robots being the most popular example \cite{desai_essential_2011, rae_framework_2015}. Typically, individuals control telepresence robots across entire office environments rather than single meetings. Recently, Dybboe et al. presented TableBot as a more focused elaboration of such an approach for hybrid meetings \cite{dybboe_tablebot_2024}. TableBot’s design innovation was to transform telepresence robots into small lightweight robots which can be placed on a meeting table and also picked up and moved by co-located participants. They explored different approaches on how such a negotiating of control happens and the benefits and drawbacks. This approach allowed more implicit interaction between remote and co-presence participants, also offering novel ways only available due to the nature of the robotic embodiment, i.e. pointing a remote participant towards something specific in the room by picking up and turning their robotic embodiment towards that. 

The possible effect of such robotic embodiments was also studied by Yasuoka et al. in what they called remote controlled avatars in physical spaces \cite{streitz_how_2023}. In contrast to the fully remote controlled TableBots, their avatar were simulated to also possess their own mind, being able to conduct autonomous actions such as waving, nodding or facing the person in the room currently speaking. They found promising results in that remote participants were more satisfied while also feeling more accepted in their presence by co-located participants.

Tackling the issue of misaligned frames of reference, the work on Hybridge tried to place remote participants with visual, tablet based embodiments around the meeting room table, while also providing remote participants the possibility to be placed in a digital twin of this room with co-located participants being replicated in there as well \cite{panda_hybridge_2024}. 

While not focusing directly on meetings, work on intent communication in the context of collaborative robotics has shown that ambient light-based approaches can integrate well into work environments and provide rich communication channels  \cite{arntz_collaborating_2022, european_foundation_for_the_improvement_of_living_and_working_conditions_humanrobot_2024}.

\subsection{Awareness and subtle communication in hybrid collaboration}

Ambient awareness systems provide a more lightweight alternative. These systems operate in the periphery of attention, providing background information about remote collaborators' status or activity without demanding focal attention. This approach builds on the concept of ``peripheral awareness,`` which enables coordination without disrupting primary tasks. It also has shown to create a feeling of relatedness among co-workers and team members \cite{hassenzahl_all_2012, Wenhart2025}.
Recent examples include WhisperChannel's audio backchannels for inclusive communication \cite{mu_whispering_2024}, which in particular addresses the need to tackle not just direct communication channels. They found that WhisperChannel helped remote and co-located participants feel more included while requiring low effort to use. However, it also created new challenges, as the easy availability of such backchannel communication means may raise the question of potential mis- or overuse in meetings. %and TeamSense's visualization of participation patterns \cite{Yas} (Yasuoka et al., 2023).

Tangible interfaces offer particularly promising directions for hybrid collaboration. Rather than attempting to recreate face-to-face interactions digitally, these approaches embed digital information in physical artifacts that can be shared across locations. Systems like LumiTouch used ambient lighting to signal presence \cite{chang_lumitouch_2001}. Such approaches align with what Hassenzahl et al. termed ``mediated intimacy`` --- using technology to create new forms of connection rather than replicating existing ones \cite{hassenzahl_all_2012}. We draw particular inspiration from research on ``suggestive communication mediums`` that express intentions through sensory experiences beyond precise information exchange \cite{Rasmussen2012}, as well as relatedness technologies, which aim to form and enhance human relationships through technological means over distance \cite{hassenzahl_all_2012, Wenhart2025}. By abstracting communication into ambient, physical signals, these approaches open new expressive channels that complement rather than replace conventional interaction modes. \break

\noindent Our work with NoticeLight aims to build upon this foundation and bridge a gap between physical direct embodiment through 1:1 mapping on robot devices such as TableBot as proposed by \cite{dybboe_tablebot_2024} and mediated, tangible communication devices, such as LumiTouch \cite{chang_lumitouch_2001}. For this, rather than viewing asymmetry as a deficiency to be overcome, we embrace Bjørn et al.'s perspective that asymmetry can be a generative design constraint \cite{bjorn_achieving_2024}. This aligns with emerging work on ``asymmetric visualization layers`` \cite{neumayr_what_2022} that give remote users unique interaction privileges that would be impossible in purely physical settings. NoticeLight occupies a unique position within this landscape. Unlike high-fidelity telepresence systems that attempt to erase boundaries between physical and digital spaces, NoticeLight acknowledges these boundaries while creating a permeable membrane between them. Unlike purely digital awareness tools, NoticeLight manifests in the physical environment of the co-located group, creating a tangible, persistent representation of remote presence. And unlike existing tangible interfaces that typically focus on one-to-one connections, NoticeLight aggregates and abstracts group states to provide awareness without overwhelming cognitive resources. By translating remote participants' digital actions into ambient environmental cues, NoticeLight creates what we term a ``peripheral presence channel`` --- a low-cognitive-load mechanism for maintaining awareness of remote participants' states and needs without disrupting the primary interaction flow. This approach aligns with the concept of ``calm technology`` \cite{weiser_designing_1996}, which moves between the periphery and center of attention as needed.

% Recent work has also explored AI-mediated approaches, using machine learning to analyze meeting dynamics and promote equitable participation. The ``Exploration of Multimodal Visual and Haptic AI-Mediation in Hybrid Meetings`` (EMVH, 2024) demonstrated how real-time feedback on speaking time distribution could encourage more balanced interactions. However, these approaches risk over-quantifying social dynamics and creating surveillance concerns.

\section{NoticeLight Concept}
\label{sec:Concept}
The NoticeLight initself is foremost envisioned as a physical, lamp-like device placed in meeting rooms where co-located participants gather, coupled with a companion application (for example as a mobile app or a plugin within a video conferencing tool) for remote participants to interact and also perceive a digital twin of the NoticeLight device. The physical device serves as a tangible proxy, translating remote participants' digital actions into ambient light patterns that provide peripheral awareness to the co-located group. Rather than demanding constant attention, NoticeLight operates at the edge of perception, becoming focal only when necessary to address specific needs but at the same time being visually perceivable for everyone in the room, thereby using a cylindrical shape. 

Our design philosophy centers on three key principles:
\begin{itemize}
    \item Abstraction over replication: Instead of attempting to recreate face-to-face cues digitally, NoticeLight abstracts and aggregates information into ambient patterns that convey essential meaning without cognitive overload.
    \item Peripheral awareness: Following concepts such as calm technology \cite{weiser_designing_1996}, suggestive communication \cite{Rasmussen2012} and relatedness technologies \cite{Wenhart2025}, NoticeLight provides awareness without demanding attention, becoming prominent only when necessary.
    \item Productive asymmetry: Rather than attempting to eliminate differences between co-located and remote experiences, NoticeLight creates new interaction possibilities that would be impossible in purely physical settings.
\end{itemize}

NoticeLight's design intentionally avoids individual attribution to prevent unintended exposure of remote participants. The system employs aggregated, anonymized representations of digital input, making signals inherently non-traceable to specific individuals. This approach aligns with Bjørn et al.'s asymmetry principles, prioritizing team cohesion over surveillance-like transparency. Consequently, the system proves most effective in established teams with existing social bonds, where collective awareness supersedes personal identification needs. In large, anonymous meetings, such abstraction could enable disruptive signaling. The NoticeLight system integrates three core functionalities designed to address specific aspects of hybrid collaboration challenges.

\begin{figure}
    \subfloat[]{\includegraphics[width=0.32\linewidth]{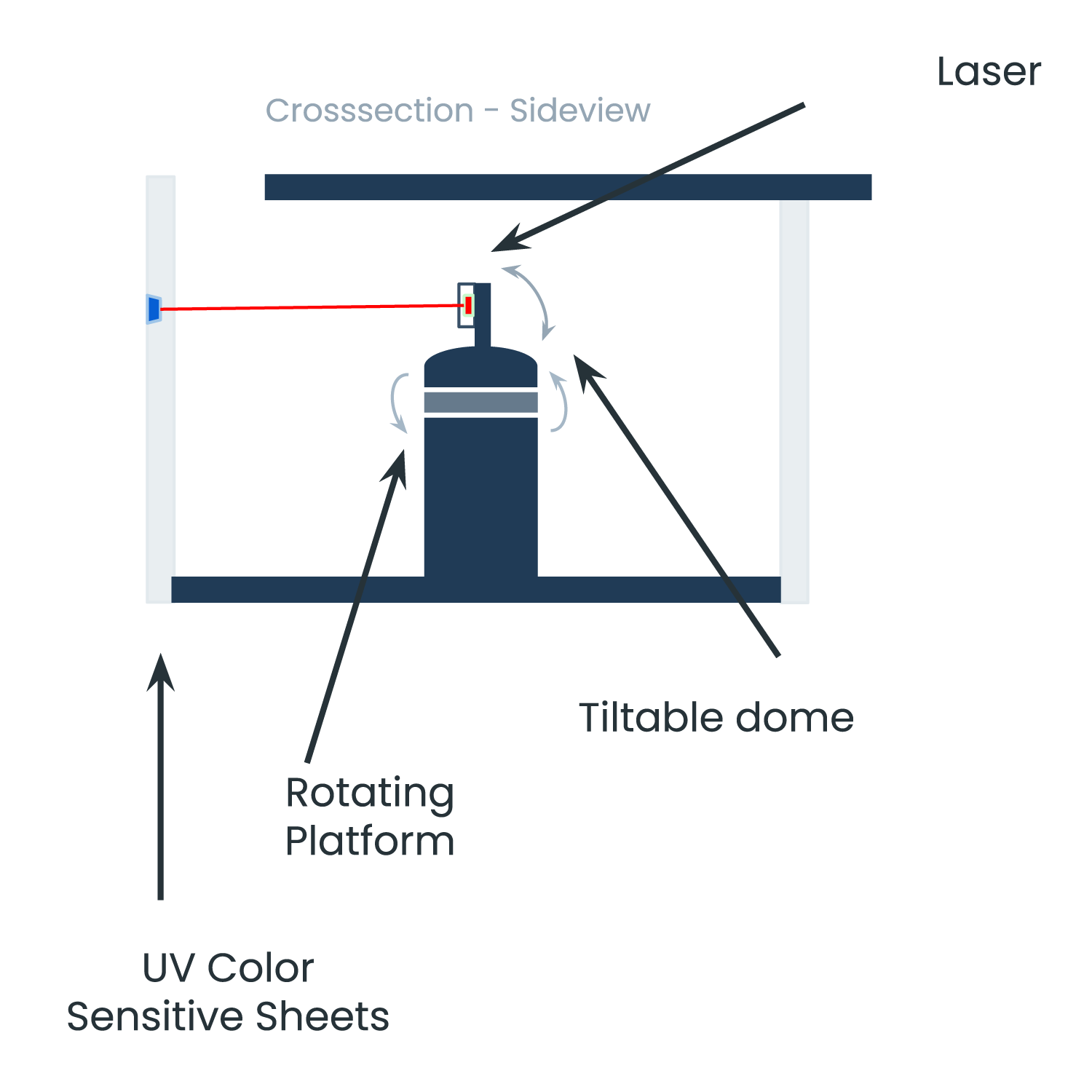}}
    \hfill
    \subfloat[]{\includegraphics[width=0.32\linewidth]{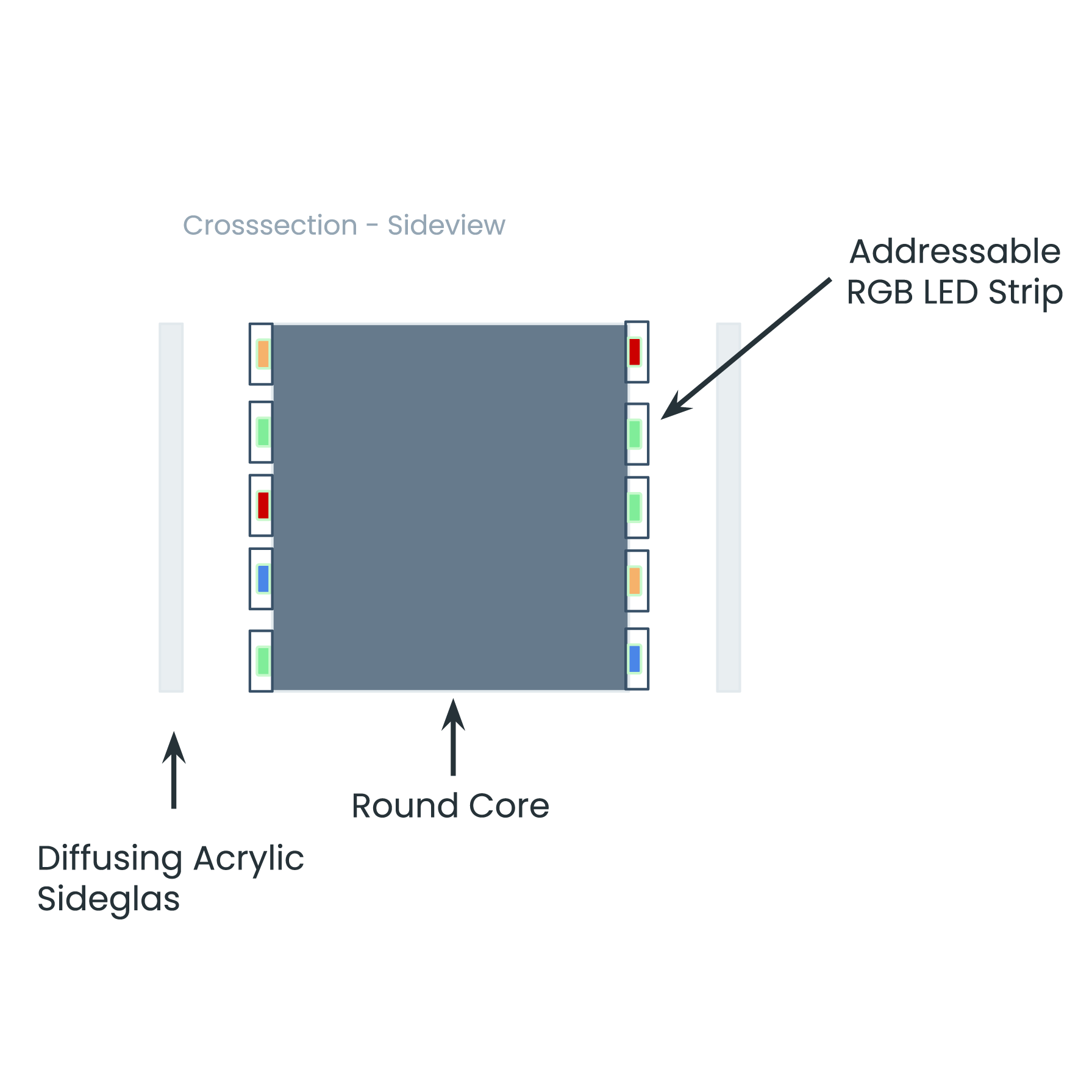}}
    \hfill
    \subfloat[]{\includegraphics[width=0.32\linewidth]{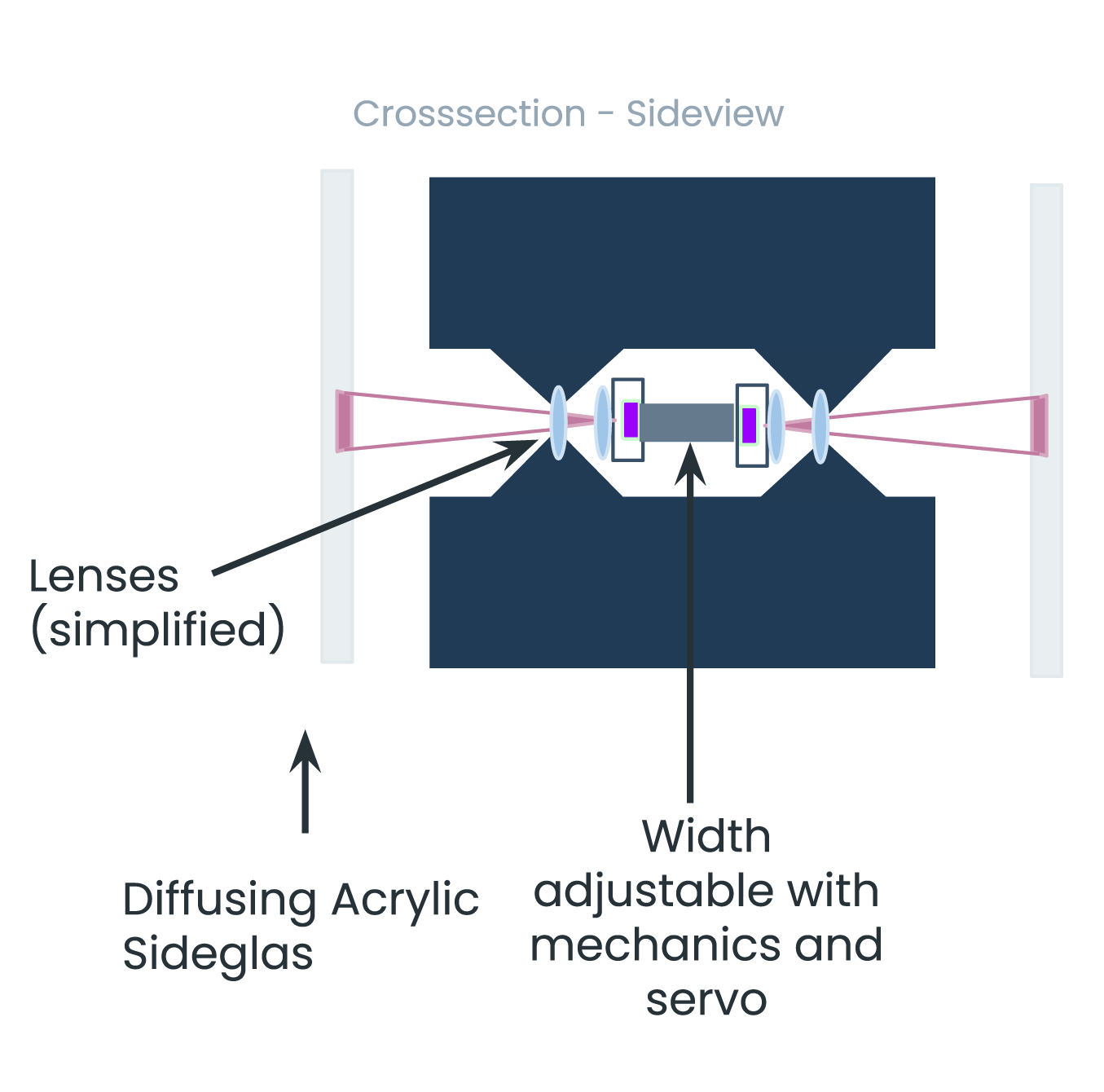}}
    \caption{a) The \textbf{Mood Dynamics} approach with a revolving light-emitting diode directed on a UV-sensitive plane, b) \textbf{Verbal Contribution Mosaic} with addressable RGB LED strip, c) \textbf{Attention Grabber} with a single RGB LED in a diffusing revolving case}
    \label{fig:NoticeLight_features}
\end{figure}

\subsection{Mood Dynamics}
Remote participants frequently lack effective backchannel communication mechanisms --- the subtle nods, smiles, and other non-verbal cues that signal agreement or engagement during presentations or discussions \cite{mu_whispering_2024, ruiz_organizing_2019}. Video streams often fail to capture these signals due to poor visibility, camera positioning, or participants disabling their cameras entirely. In addition, the absence of camera visibility can lead to the perspective that remote participants are not attentive or contributing \cite{Saatci2019}. Research has shown that visualizing emotions can be both helpful and potentially distracting \cite{kurosu_emotion_2023}. At the same time, happy emotional expressions support the feeling of co-presence \cite{ahmed_reach_2016}. NoticeLight navigates this tension by providing an abstracted representation of emotional feedback that is visible yet non-intrusive. Camarillo-Abad et al. note that technologies can both support existing non-verbal communication and create entirely new forms --- NoticeLight takes the latter approach, transforming digital input into novel visual expression \cite{ruiz_organizing_2019}.

\textbf{Concept}: The approach taken by NoticeLight allows remote participants to explicitly express agreement, for example by a button in the companion app. However, this is then aggregated and converted to a dynamic wave pattern which, similar to a lifeline ECG visualization for a patient in a hospital. It provides the co-presence audience with an overall visualization of the level of agreement and engagement of remote participants. A flat line is to be considered as neutral here, while one or more agreement expressions will accordingly lead to harmonic sine-wave. The percentage of the group that communicates agreement is represented in the amplitude of the wave. The visual effect of the wave integrates a gentle pulsing or flowing light, which can also be coupled to the number of participants signaling agreement. This pattern then gradually decays over time through an exponential decay function, providing temporal context without creating persistent distraction. 
From a technical standpoint, we aim to realize this linegraph by a revolving light-emitting diode on a UV-sensitive plane, similar to the concept used in CRT televisions and oscilloscopes (see Fig.\ref{fig:NoticeLight_features}a)).
We purposefully decided not to allow the communication of disagreement here, as we think that in that case the anonymous nature of the communication could lead to misuse and create new ways of cyber mobbing. To be noted is that this kind of communication mechanism may allow individuals to signal support or agreement even for individual statements, but the reception of the visualization will be different and not allow that trace back on that level of granularity. Again, this is on purpose, as we want to communicate moods, and use the possibility of stating agreement as proxy.

\subsection{Verbal Contribution Mosaic}
Unbalanced speaking time is a persistent challenge in hybrid meetings, with co-located participants typically dominating conversational flow. Kuzminykh and Rintel observed that remote participation in work meetings is generally associated with lower motivation to engage both behaviorally and cognitively, creating a self-reinforcing cycle of non-participation \cite{kuzminykh_low_2020}, while others have observed dynamics that lead to a primary room dominance \cite{Saatci2019}. This asymmetry remains, however, intransparent and difficult to address without, for example, singling out a remote participant and ask them to speak – which of course can also be very uncomfortable and stressful for that said participant. Similarly, explicitly displaying speaking time metrics for each participant could potentially create social pressure or even surveillance concerns.  Our goal with NoticeLight is more subtle, i.e. we aim to provide a certain level of awareness regarding the distribution and dynamics of speaker allocation. NoticeLight approaches this challenge through abstract visualization that provides awareness without singling out individuals. We therefore provide real-time feedback on interaction dynamics which help to promote more balanced participation in meeting types and situations where this is desirable.

\textbf{Concept}: NoticeLight includes a microphone array which applies voice recognition and diarisation algorithms to identify different speakers and track their contributions \cite{bai_speaker_2021}. Thereby, we can attribute speaking time for each participant, both co-located and remote. As we do not wish to identify individuals, no additional information is required. This data is transformed into a dynamic mosaic of colored light segments on the device, where each color or shade of color represents an individual. The latter helps to maintain anonymity in smaller meetings, where otherwise a careful observer of the Mosaic could figure out which color represents which person. The proportional vertical size of each colored segment reflects the relative distribution of speaking time, while every turn-taking creates a new segment on the x-Axis. The mosaic updates periodically (every 15-30 seconds) based on a moving window of recent activity, providing an evolving picture of participation patterns without demanding constant attention and in addition also making it more difficult to identify the relationship between colors and participants.

Technically, this can be realized by wrapping an addressable RGB LED stripe around a the NoticeLight cone in multiple rows, and lighting up the corresponding amount of LEDs in the specific color of the participant or group of participants (see Fig.\ref{fig:NoticeLight_features}b)). Visualizing this on a small area and repeat this pattern ensures that this information can be perceveid from different viewpoints on the NoticeLight.

% Unbalanced speaking time is a persistent challenge in hybrid meetings, with co-located participants typically dominating conversational flow. Kuzminykh and Rintel observed that ``remote participation in work meetings is generally associated with lower motivation to engage both behaviorally and cognitively,`` creating a self-reinforcing cycle of non-participation \cite{kuzminykh_low_2020}.

% While explicitly displaying speaking time metrics could potentially create social pressure or even surveillance concerns, real-time feedback on interaction dynamics can promote more balanced participation when implemented thoughtfully. NoticeLight approaches this challenge through abstract visualization that provides awareness without singling out individuals.

% Interaction Design: The system uses current voice recognition and diarisation algorithms to identify different speakers and track their contributions \cite{bai_speaker_2021}. This data is transformed into a dynamic mosaic of colored light segments on the device, where each color represents either an individual (in smaller meetings) or a group (in larger meetings). The proportional size of each colored segment reflects the relative distribution of speaking time.

% The mosaic updates periodically (every 45 seconds) based on a moving window of recent activity, providing an evolving picture of participation patterns without demanding constant attention. Unlike direct speaker identification, this abstracted representation maintains a degree of anonymity while still conveying essential information about conversational balance.

\subsection{Attention Grabbing}
Remote participants often struggle to interject in conversations or have their digital signals (raised hands, chat messages) noticed by the co-located group. Mostly, this is not on purpose but due to the technical infrastructure and setup \cite{nakanishi_hybrid_2019}, where such cues cannot be seen by everyone or are completely hidden during certain parts of a meeting (e.g. during screensharing). This contributes to what Bjørn et al. describe as the increased ``articulation work`` required in hybrid settings --- the additional effort needed to coordinate interaction across different spatial configurations \cite{bjorn_achieving_2024}.
Research on asymmetric attention levels highlights how technical infrastructure shapes where and how attention is directed in hybrid meetings \cite{kuzminykh_classification_2020}. In addition, the physical separation between remote and co-located participants often leads to attention being either on one or the other group, but not both at the same time  \cite{saatci_reconfiguring_2020}.

\textbf{Concept}: NoticeLight addresses this by creating a persistent, physical representation of remote participants’ attempts to contribute which remains visible until addressed. Remote participants can trigger an ``attention needed`` signal either manually (by clicking a button in the companion app) or automatically (when a raised hand in the video conferencing platform has been active beyond a threshold duration). This signal appears as a distinct light pattern on the NoticeLight device --- for example, which visually encodes both the number of participants who made this request and the time that has passed since. Once a participant has spoken, the attention signal fades away. 
This creates a persistent but non-disruptive reminder, that someone wishes to contribute, increasing the likelihood that remote participants will be included in the conversation without requiring them to verbally interrupt. The system does not replace a video-conferencing tool to negotiate who may speak first or who it is that wants to contribute. Instead, we focus on the essential aspect that the visual presentation on NoticeLight shares the burden of recognizing the remote participants’ willingness to contribute to all co-located participants, not just the person operating the remote video conferencing tool. 
Technically, we aim to realize this with a single RGB LED in a diffusing revolving case (see Fig.\ref{fig:NoticeLight_features}c)). The LED flashes shortly and repeatedly, if a participant expresses an attention grab. The breaks between the flashes decrease, the more time goes by. If multiple participants perform an attention grab, the LED flashes in different colors for the different participants.  Alternatively, in case the flashing turns out to be too disruptive, we suggest a constant LED light that gradually increases in intensity over time and grows in widths based on the number of participants trying to grab attention.

\subsection{Discussion}
As NoticeLight remains at a conceptual stage, we recognize that its anticipated benefits are speculative and open to challenge. Throughout our design process, we have explored multiple variants, weighing their respective advantages and drawbacks. Achieving an optimal configuration will require a nuanced understanding of collaborative styles and specific meeting tasks, which we plan to address through co-design workshops and observational studies within our PRAESCO research project, using NoticeLight as a probe to test our assumptions.

Some notable technical and conceptual risks of our current concept include:
\begin{itemize}
    \item Voice diarization: While algorithms are robust, dynamic meetings with cross-talk and noise may pose challenges.
    \item Anonymity: The Verbal Contribution Mosaic’s color-coding could enable de-anonymization in small teams, so we will explore non-persistent encodings.
    \item Disagreement signaling: Excluding dissent to prevent misuse may silence minority views; alternative abstractions for dissent will be considered.
    \item Subtle cues: Calm visual signals may go unnoticed in heated discussions, so adaptive mechanisms will be explored.
    \item Monolithic Design: NoticeLight could also be modular, with features tailored to meeting types, extending beyond the presented three core areas.
\end{itemize}

Our technical concepts, as illustrated in Fig \ref{fig:NoticeLight_features}, are first iterations; other modalities and realizations are possible. Designers must ensure output modalities are usable in real meeting contexts --- e.g., light intensity may be affected by ambient conditions, suggesting alternatives like pulsating or frequency-modulated signals. We excluded auditory feedback to avoid disruption but are considering individualized haptic feedback via wearables linking remote participants with co-located sponsors.

% Regarding the technical realization, we aimed to provide a first set of specific ideas, as seen in Fig \ref{fig:NoticeLight_features}. We are aware that some aspects could also be realized in various other ways and also other modalities could be considered. Beyond ensuring compatibility between input data sources and output expressions, designers must evaluate how effectively a chosen modality and its parameters align with usability requirements in real-world meeting contexts. For instance, while light intensity could theoretically encode continuous values, practical limitations like ambient lighting interference or low perceptual resolution for subtle brightness variations may compromise effectiveness. These constraints highlight the need to investigate alternative visualization approaches, such as pulsating patterns with frequency modulation, to convey gradient information more reliably. We deliberately chose not to incorporate auditory feedback, as we believe it would be too disruptive in a meeting setting. Nonetheless, we are exploring the possibility of providing individualized haptic feedback by connecting remote participants with designated co-located “sponsors” through wearable devices.

NoticeLight intentionally prioritizes empowering remote participants through asymmetry inversion. The Attention Grabbing module exemplifies this by granting remote members exclusive access to persistent physical signals that co-located participants cannot employ, thereby rebalancing conversational agency in turn-taking initiation. This strategic exclusion of co-located participants from requiring additional interfaces (e.g., apps for mood signaling) preserves the natural interaction rhythms of physical meetings. An exception exists for the Verbal Contribution Mosaic, which passively includes all participants through existing speaking behaviors without intervention. We intentionally avoid solutions that dilute established co-located dynamics by forcing partial or full digital immersion, as such approaches risk eroding the very benefits of physical collaboration that hybrid configurations aim to preserve.

The potential of NoticeLight extends significantly to people with disabilities, who often face unique barriers in both traditional and hybrid work environments. By providing a tangible, ambient interface for remote participation, NoticeLight can help address issues of visibility, inclusion, and equitable contribution that are particularly acute for employees with disabilities. For example, the ability to signal engagement, mood, or the need to contribute through accessible digital interfaces can enhance opportunities for participation in meetings for individuals navigating mobility, sensory, or cognitive barriers, by reducing reliance on verbal or physical cues. Furthermore, NoticeLight’s abstraction and peripheral awareness mechanisms can reduce cognitive load and social anxiety, offering a less intrusive way for team members with disabilities to remain present and involved. As hybrid work continues to evolve, such robotic embodiments hold promise for making workplaces not only more inclusive but also more adaptable to diverse needs, supporting the broader goal of disability-inclusive employment.

\section{Conclusion}
In summary, the NoticeLight concept showcases how embracing socio-technical asymmetries in hybrid collaboration can lead to more inclusive and dynamic workplace interactions. By translating remote participants’ digital presence into tangible, ambient cues, NoticeLight supports peripheral awareness and fosters equitable participation without overwhelming users. This conceptual approach positions robotic interfaces not as replacements for human presence, but as mediators that enhance group synergy and well-being. Looking ahead, the integration of such robotic embodiments signals a shift toward workplaces where robots actively shape social dynamics, balance participation, and adapt to evolving human needs. As human-robot synergy deepens, future workplaces will benefit from robots that are not only technically competent but also socially attuned, paving the way for richer, more responsive collaborative environments.

The NoticeLight concept is part of a larger joint research project PRAESCO \footnote{https://ihri.reha.tu-dortmund.de/research/projects/praesco/}, which will allow us to continue to work on that for several years, including field studies over days and weeks in various companies across Europe. Therefore, we aim to both technically realize NoticeLight in various iterations and evaluate its effectiveness in hybrid collaboration settings. 

\begin{acknowledgments}
As part of PRAESCO, this work is funded by the German Federal Ministry of Research, Technology and Space (BMFTR) and the European Social Fund (ESF Plus) under the “Future of Work” program (FKZ: \href{https://foerderportal.bund.de/foekat/jsp/SucheAction.do?actionMode=view&fkz=02L23B080}{02L23B080}) and managed by the Project Management Agency Karlsruhe (PTKA). The authors take full responsibility for its content. We would like to thank our students Christopher Bussick, Leona Dunke, Judith Humkamp, Svetlana Kuhn, and Nicola Pache for their early work on this topic and for coining the term ''NoticeLight''.

\end{acknowledgments}
  % The research is supported by the Federal Ministry of Education and Research of the Federal Republic of Germany under the funding code FKZ: \href{https://foerderportal.bund.de/foekat/jsp/SucheAction.do?actionMode=view&fkz=16KIS1631}{16KIS1631}. The responsibility for the content of this publication lies with the authors.

  \section*{Use of Generative AI}
  During the preparation of this work, the author(s) used Perplexity with various models (Gemini Pro 2.5, GPT4.1, Claude 3.7) in order to: drafting content, improve writing style, grammar and spelling check, paraphrase and reword, and peer review simulation. After using this tool/service, the authors thoroughly reviewed and edited the content and take full responsibility for the publication’s content.
  
  %Generative AI was used to improve the quality of existing text written by the authors. The authors acknowledge that the resulting work in its totality is an accurate representation of the authors’ underlying work and novel intellectual contributions and is not primarily the result of the tool’s generative capabilities. The authors accept responsibility for the veracity and correctness of all material in their work, including any computer-generated material according to the ACM policy on authorship\footnote{ACM Policy on Authorship: \url{https://www.acm.org/publications/policies/frequently-asked-questions}}.

%% Define the bibliography file to be used
\bibliography{MS}

% \appendix


% The sources for the ceur-art style are available via
% \begin{itemize}
% \item \href{https://github.com/yamadharma/ceurart}{GitHub},
% % \item \href{https://www.overleaf.com/project/5e76702c4acae70001d3bc87}{Overleaf},
% \item
%   \href{https://www.overleaf.com/latex/templates/template-for-submissions-to-ceur-workshop-proceedings-ceur-ws-dot-org/pkfscdkgkhcq}{Overleaf
%     template}.
% \end{itemize}

\end{document}